# Particulate Matter Exposure at a Densely Populated Urban Traffic Intersection and Crosswalk

Hong-di He[a, *], H. Oliver Gao[b,c]

[a] Center for ITS and UAV Applications Research, School of Naval Architecture, Ocean & Civil Engineering, Shanghai Jiao Tong University, Shanghai 200240, China

[b] School of Civil and Environmental Engineering, Cornell University, Ithaca, NY 14853, USA

[c] Center for Transportation, Environment, and Community Health, Cornell University, Ithaca, NY 14853, USA

(Corresponding author: *hongdihe@sjtu.edu.cn)

## Graphic Abstract

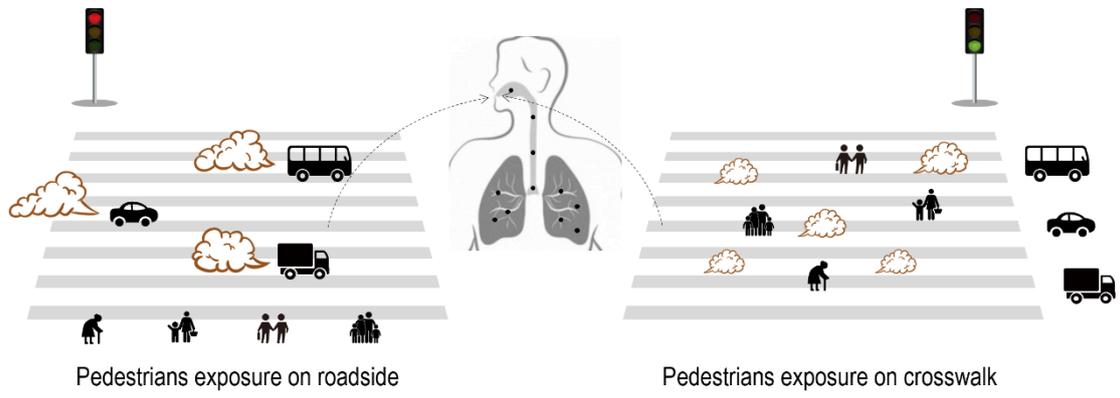

Pedestrians exposure on roadside          Pedestrians exposure on crosswalk

# Abstract

Exposure to elevated particulate matter (PM) pollution is of great concern to both the general public and air quality management agencies. At urban traffic intersections, for example, pedestrians are often at a higher risk of exposure to near-source PM pollution from traffic while waiting on the roadside or while walking in the crosswalk. This study offers an in-depth investigation of pedestrian exposure to PM pollution at an urban traffic intersection. Fixed-site measurements near an urban intersection were conducted to examine the variations in particles of various sizes through traffic signal cycles. This process aids in the identification of major PM dispersion patterns on the roadside. In addition, mobile measurements of pedestrian exposure to PM were conducted across six time intervals that correspond to different segments of a pedestrian's journey when passing through the intersection. Measurement results are used to estimate and compare the cumulative deposited doses of PM by size categories and journey segments for pedestrians at an intersection. Furthermore, comparisons of pedestrian exposure to PM on a sunny day and a cloudy day were analyzed. The results indicate the importance of reducing PM pollution at intersections and provide policymakers with a foundation for possible measures to reduce pedestrian PM exposure at urban traffic intersections.

**Keywords:** Pedestrian exposure; Particulate matter; Crosswalk; Intersection;

## 1 Introduction

Air pollution is a worldwide problem affecting both the environment and human health (Fann et al., 2012; Matus et al., 2012; Huang et al., 2014; Landrigan et al., 2019; Pan et al., 2019; Sadeghi et al., 2020). As one of the most common air pollutants, particulate matter (PM) can be deposited in the thoracic area of the respiratory system, causing serious health problems such as birth and developmental defects, premature mortality, cardiovascular problems, and cancer (Brook et al., 2010; Heo et al., 2017; Scungio et al., 2018; Khan et al., 2019). Traffic-related particles, especially in densely populated urban areas, have been identified as one of the major sources of PM pollution, posing a significant threat to people in and near the traffic (Wang and Gao, 2011; Quiros et al., 2013; Adeniran et al., 2017). Consequently, pedestrian exposure to PM pollution in various traffic microenvironments has become an important area of study for researchers and regulators alike (Goel and Kumar, 2016; Wang et al., 2018).

It is not difficult to recognize that traffic intersections constitute hotspots for air pollution with elevated airborne particles. At an urban traffic intersection, vehicles frequently stop with their engines idling during the red-light period, before accelerating (with intensified emissions) during the green-light period. Over the past few decades, many scholars have studied relevant issues related to this topic. Gokhale and Raokhande (2008) reported that changes in driving patterns (i.e., idle, acceleration, deceleration, and cruising patterns) result in incomplete combustion within the engine, leading to an increased release of PM from the tailpipe. Lu and He (2012) observed periodic variations in PM pollution that correlate with traffic signal changes at urban traffic intersections. Kumar and Goel (2016) examined pedestrian exposure to $PM_{10}$, $PM_{2.5}$, and $PM_1$ (particulate matter with aerodynamic diameters smaller than 10, 2.5 and 1 μm respectively) on a roadside by estimating the total respiratory deposition dose (RDD) from data collection at traffic intersections. They found that at a 4-way traffic intersection, pedestrian exposure to $PM_{10}$ and $PM_{2.5}$ under a congested-flow condition was 7.3 and 1.2 times higher than that under a free-flow condition. This work advances the understanding of roadside exposure to PM at intersections and assists in making an informed choice to limit exposure at such pollution hotspots. Adeniran et al. (2017), meanwhile, measured the total suspended particles (TSP), $PM_{10}$, $PM_{2.5}$, and $PM_1$, at intersections across multiple seasons, finding that pedestrian exposure on dry days is more serious than that on wet days in Ilorin, Nigeria.

Existing studies provide different perspectives for understanding the variations in PM pollution at urban traffic intersections. However, the key issue of pedestrian exposure to PM at intersections, especially in crosswalks, has seldom been the focus of those studies. As rightly argued by some researchers, a pedestrian is inevitably exposed to significantly higher PM concentrations while walking in a crosswalk during the green traffic light period (De Nazelle et al., 2009; Ishaque et al., 2018). It is therefore valuable to carry out an in-depth investigation of pedestrian exposure at intersections both on the roadside and in the crosswalk. Such studies are particularly rare, and remain very much needed for cities in China.

This study aims to addresses the following questions related to pedestrian exposure to PM pollution at intersections: (i) How does the traffic signal cycle affect the variation of particles of various sizes on roadsides at a typical 4-way traffic intersection? (ii) What is the dispersion pattern of different particles at an urban intersection? (iii) How do the concentrations and exposure levels vary during different pedestrian journeys in the crosswalk? (iv)What is the difference in pedestrian exposure to particles in crosswalks while crossing and on the roadside while waiting?

## 2 Material and methods

### 2.1 Sampling location

The sampling area was located in Shanghai, China. With a population of 24.28 million as of 2019, Shanghai is the most populous urban area in China and the second most populous city in the world. The high population density of this city is serviced by an expansive expressway network including the Inner Ring Road, the North-South Road, and Yan'an Elevated Road. According to the Shanghai Master Plan for 2017–2035, there will be 320 km of viaducts built in this city in the future, accounting for 25% of the total main road length (Hao et al., 2019).

The PM pollution measurements were performed at a selected traffic intersection along Gonghe Road and Guangzhong Road in the central area of the JinAn district, Shanghai (Fig. 1 (a), (b), and (c)). The Gonghe Road is a dual carriageway with five 5-lanes in each direction, which is 40 m wide in total. In the center of the carriageways, there exists a green infrastructure barrier of 6 m. In addition, the North–South Elevated Road is parallel to the GongHe road and is located above it. The North–South Elevated Road is a dual carriageway with four lanes in each direction, and is 30 m wide in total. Therefore, this selection of the sampling location is considered as a typical case in Shanghai, in order to make the research more impactful (Hao et al., 2019).

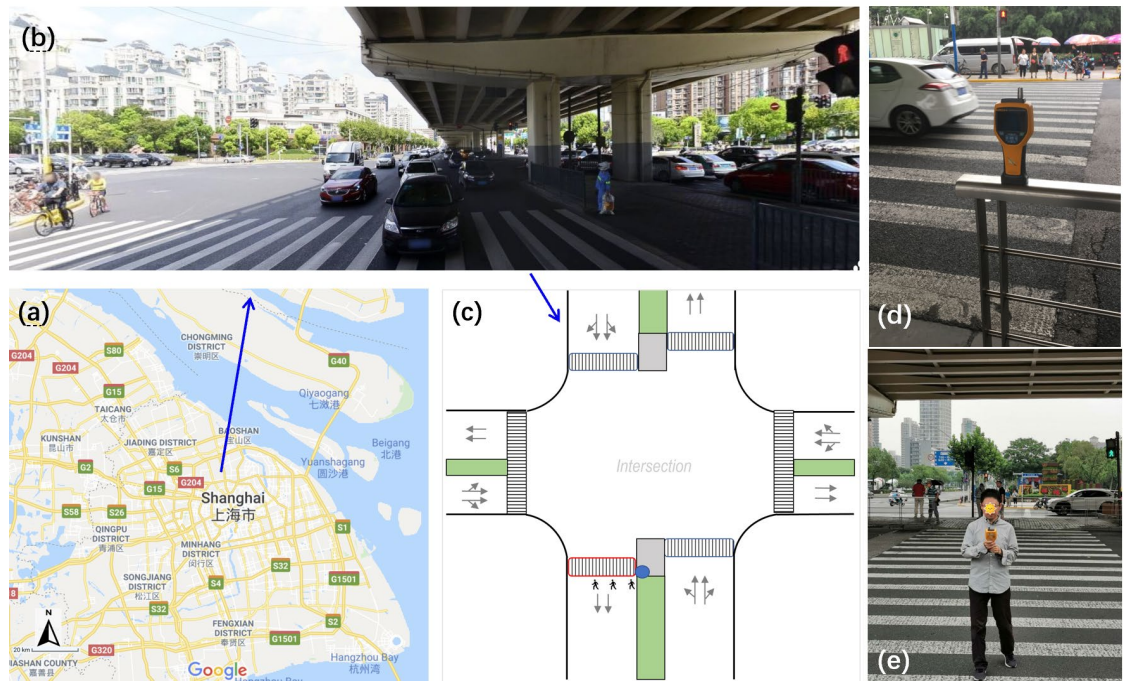

Fig. 1 The sampling condition. a) the sampling area (the source is from Google Map); b) the sampling location (the source is from Baidu Map); c) the configuration of measured intersection; d) the fixed measurement on roadside; e) the mobile measurement on crosswalk.

The selected intersection has the configurations of four traffic phases, as shown in Fig. 2 (a). In every traffic phase, the vehicles are allowed to cross the intersection under their own green signal at a particular time. During their red signal period, pedestrians are allowed to pass through the crosswalk. Such a system works to avoid traffic crashes from different directions, guaranteeing that pedestrians can cross the road safely. This type of phasing is representative and has been widely adopted in urban areas (Kumar and Goel, 2016).

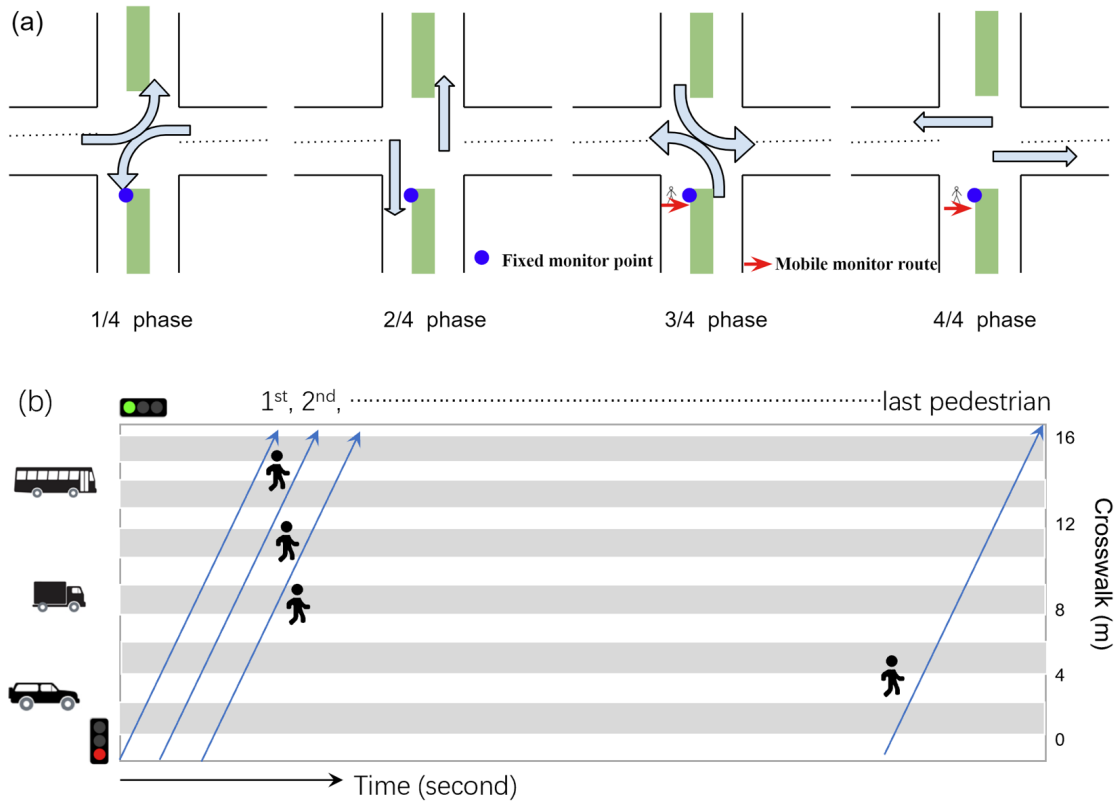

Fig. 2 Sketch of traffic phase at the selected urban traffic intersection (a) and varied departure route of pedestrians on crosswalk (b)

## 2.2 Sampling equipment

The measurement of particulate matter was carried out using a Fluke 985 Particle Counter (Fluke Corporation, USA), which can detect particles every second by using a laser diffraction technique for size differentiation from submicron to millimeter (Fig. 1 (d)). Using this instrument, the number concentration of particles larger than 0.3 $\mu$m was obtained and classified into six categories: 0.3-0.49 $\mu$m, 0.5-0.99 $\mu$m, 1-1.99 $\mu$m, 2-4.99 $\mu$m, 5-9.99 $\mu$m, and $\geq 10$ $\mu$m. In previous studies (Hinds, 1999; Kumar and Goel, 2016; Liu et al., 2020), the mass concentration of the particles is usually used to assess human exposure to particles. Hence, in order to determine the pedestrian exposure to PM in crosswalks in the same manner, a simple formula has been adopted to convert the particle number concentration to the particle mass concentration in this study. The conversion formula has been widely applied by many scholars and validated to provide a bridge between particle number concentration and mass concentration (Tuch et al., 2000; Wittmaack, 2002; Tittarelli et al., 2008; Araji et al., 2016; Zhai and Albritton, 2020). It is defined as follows:

$$C_i = N_i \times \rho \times \frac{4}{3} \times \pi \times (\frac{D_p^i}{2})^3 \qquad i = 0.3 - 0.49, \ 0.5 - 0.99, \ 1 - 1.99, \ 2 - 4.99, 5 - 9.99$$

(1)

where $C_i$ represents the particle mass concentration ($\mu$g/cm$^3$) in each category and $N_i$ is the corresponding number concentration (particles/cm$^3$). $D_p^i$ is the particle diameter in every category and $\rho$ is the particle density. As suggested by many references (Tittarelli et al., 2008; Araji et al., 2016; Zhai and Albritton, 2020), the algorithm used to transform particle numbers to mass assumes particles are spherical and have a density of 1.65 g/cm$^3$. For each category, the mean dimeter is obtained, that is, $D_p^{0.3-0.49} = 0.4 \ \mu m$, $D_p^{0.5-0.99} = 0.75 \ \mu m$, $D_p^{1-1.99} = 1.5 \ \mu m$, $D_p^{2-4.99} = 3.5 \ \mu m$, $D_p^{5-9.99} = 7.5 \ \mu m$ (Araji et al., 2016; Zhai and Albritton, 2020). In this study, the mass concentration of PM$_1$ (C$_{0.3-0.49}$ and C$_{0.5-0.99}$) was calculated to reflect the submicron particles, while PM$_{10}$ (C$_{0.3-0.49}$, C$_{0.5-0.99}$, C$_{1-1.99}$, C$_{2-4.99}$ C$_{5-9.99}$) was calculated to reflect the coarse particles. It should be noted that due to the categories measured using this equipment, PM$_2$ (C$_{0.3-0.49}$, C$_{0.5-0.99}$, C$_{1-1.99}$) rather than PM$_{2.5}$ was obtained to approximate the variation of fine particles.

## 2.3 Fixed-site measurement on the roadside of intersection

The fixed-site measurement was conducted on the roadside of the chosen intersection. The equipment was placed on one side of the road (Fig. 1 (d)). In contrast to the opposite side, the selected side of the road is located under the elevated road, in which pollutants are not easily dispersed (Hao et al., 2019). In addition, the PM concentrations on the opposite side of the road are adjacent to the pedestrian lane, in which the pollutants are more influenced by many random factors such as moving pedestrians and right-turning vehicles from the perpendicular road (Tiwary et al., 2011, Wang et al., 2018). Hence, the current side of the road was selected in this study.

During the green-light period, vehicles are allowed to cross the intersection, producing more pollutants. The fixed-site measurement in this period can be used to capture the pollution level, revealing the extent of pedestrian exposure while waiting on the roadside. During the red-light period, vehicles are forbidden from moving through the intersection. During this period, fixed-site measurements can be used to identify the dispersion characteristics of traffic-induced pollutants during the green-light period. Therefore, fixed-site measurements were conducted on the roadside of the urban intersection for a whole traffic signal period.

## 2.4 Mobile measurements on the crosswalk of intersection

When the traffic signal changes from green to red, vehicles are forbidden from moving and the waiting pedestrians will begin to pass through the crosswalk (Fig. 2(b)). Because of their random arrival, pedestrians walk on the crosswalk at various departure times. This implies that the pedestrians are exposed to different PM concentrations across varied intervals that correspond to different segments of a pedestrian's journey when passing through the intersection (Fig. 1(e)). To gain a better comparison, the mobile measurements were conducted at six departure times (0 s, 5 s, 10 s, 15 s, 20 s, and 30 s). It should be noted that the red-light period allowing pedestrians to cross the road is 80 s. This means that each of the six departure journeys listed above allows pedestrians to pass through the crosswalk successfully.

The portable equipment was lifted up by hand, and the measured height was about 1.3 m, which is close to the breathing height for adults (~1.6 m) and children (~1 m). The equipment was moved at a speed of 1.5 m/s, which is the mean speed at which pedestrians use the crosswalk (Li et al., 2005; Marisamynathan and Vedagiri, 2015; Guo et al., 2019). Such an arrangement can efficiently trace pedestrian exposure during their entire journey when using the crosswalk. In order to eliminate a random effect, the measurements of each journey were carried out over ten repetitions, with the mean values then used for investigation in this paper.

## 2.5 Data collection

In previous studies, most scholars used multi-day data to explore the daily variation of PM in a city area (Tiwary et al., 2011). This can eliminate random factors and capture stable characteristics. In this study, we use multi-period data to explore the PM variation within a traffic signal period. Hence, the multi-period data for fixed-site measurements and mobile measurements were collected. Additionally, in order to identify the influence of the weather, the measurements were performed at evening rush hour (4:00-7:00 pm) on a sunny day (Dec. 1, 2018) and a cloudy day (October 23, 2018). On the sunny day, the windspeed was 4.65 m/s and the relative humidity was 48.1%. On the cloudy day, the wind speed was 2.73 m/s and the relative humidity was 60%. Both wind directions were blowing from the north.

In the meantime, the traffic volume at the intersection was recorded. According to the video record during the measurement, the mean traffic volume from the north direction on Gonghe Road was about 115 vehicles per 215 s (about 1925 vehicles per hour) during the green period on the sunny day. The corresponding traffic volume was 122 vehicles per 215 seconds (approximately 2026 vehicles per hour) on the cloudy day. On both days, approximately 9% of the vehicles were buses and trucks, which mostly run on diesel oil and emit more pollutants (Wang et al., 2018; Khan et al., 2019). Obviously, it is one of the busiest traffic intersections in Shanghai (Hao et al., 2019).

## 2.6 Estimation of pedestrian exposure to PM at intersection

To quantify the health impact of particle inhalation at these levels, the respiratory deposition dose (RDD) is utilized to assess the health risk of pedestrian exposure to PM at intersections. RDD is an integrated exposure-risk function, which is dependent on exposure concentration, the deposition fraction ($DF$), and the duration time ($T$) spent in each activity as well as breathing frequency and tidal volume for the different activities. RDD is regarded as an important indicator to assess the health risk of human exposure to PM and has been adopted in many studies (Hinds, 1999; Kumar and Goel, 2016; Qiu et al., 2018; Liu et al., 2020). The RDD is defined as follows:

$$\text{RDD of } PM_i = V_T \times f \times \text{DF}_i \times \text{PM}_i \times T \quad i = 1, 2, 10 \qquad (2)$$

$$\text{DF} = \text{IF} \times (0.058 + \frac{0.911}{1+\exp{(4.77+lnd_p)}} + \frac{0.943}{1+\exp{(0.508-2.58lnd_p)}}) \qquad (3)$$

$$\text{IF} = (1 - 0.5(1 - \frac{1}{1+0.00076d_p^{2.8}}) \qquad (4)$$

where $V_T$ is the tidal volume within the exposure duration $T$, $f$ is the frequency of breathing, and $DF_i$ and $PM_i$ represent the deposition fraction and mass concentration, respectively, for particle $i$.

According to existing studies (WHO, 2006; Guo et al., 2020), children are more vulnerable to PM even at low values. Consequently, in this study, both child exposure and adult exposure were also considered. The tidal volume for adults was 1250 $cm^3$ per breath, while for children it was 583 $cm^3$ per breath (Sanchez-Soberon et al., 2015). The breath frequency for adults is 20 times per minute, compared to 32 for children (Sanchez-Soberon et al., 2015). On the basis of these parameters, the estimations of $DF_i$ for PM$_1$, PM$_2$, and PM$_{10}$ were determined, and the corresponding RDDs were obtained.

# 3 Results and discussion

## 3.1 Variations of PM at the roadside of intersection through fixed-site measurements

As expected, the original data exhibits periodic variations corresponding to traffic signal intervals. The mean variation of PM within a traffic signal period on a sunny day is shown in Fig. 3. In the first phase, vehicles approaching from the right start to move and the concentration of all sizes of particles tended to increase. In the second phase, the vehicles in the straight lane began to move, and the corresponding concentrations continued to increase. Evidently because of the increased traffic volume, the PM concentration in this phase was higher than during the first phase. In the third and fourth phases, the vehicles were forbidden from moving, which indicates that the accumulated PM in the former two phases began to diffuse and finally decreased to a relatively stable value. This observation is in agreement with a previous study (Tiwary et al., 2011). The box-plots corresponding to Fig. 3 are shown in Fig. S1 (Supplementary Information).

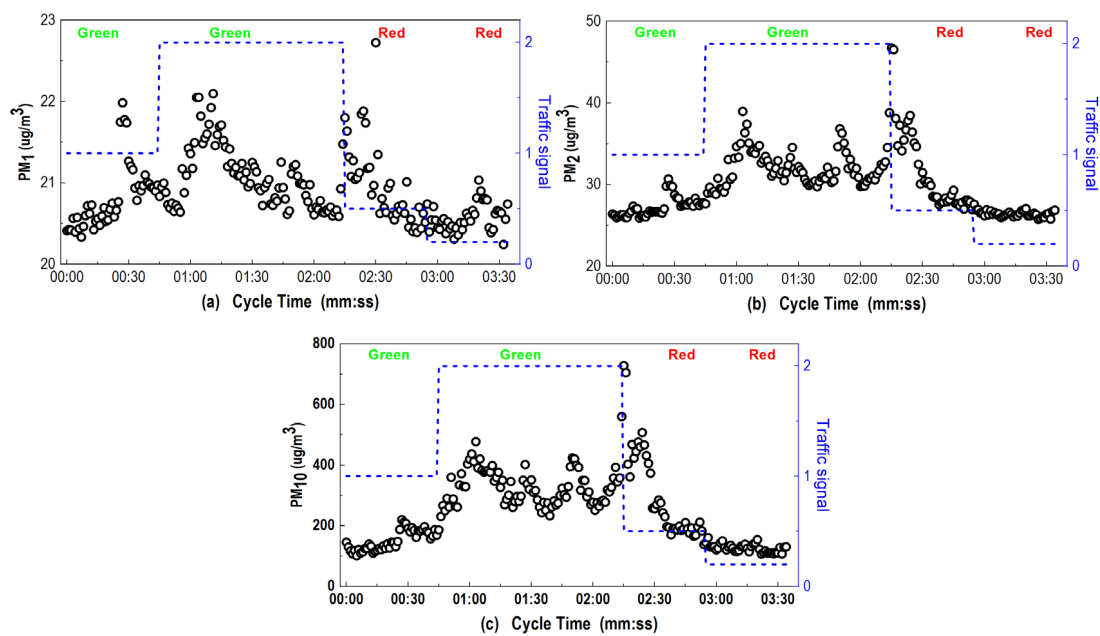

Fig. 3 Mean variation of $PM_1$, $PM_2$, and $PM_{10}$ during the whole cycle time on roadside measurement on the sunny day (Traffic index: 1-green traffic light period for left turning vehicles on perpendicular road, 2-green traffic light period for straight vehicles, 0.5-red traffic light period for straight on perpendicular road, 0.2 -red traffic light period for left turning on perpendicular road).

Additionally, we found that when the traffic lights switch from green to red, the PM concentrations for all particle sizes continued to rise for a short period, rather than decreasing instantaneously. There may be two reasons for this. First, at the end of the green light period, the vehicles in the queue have covered a long distance and run at high speeds. Second, in order to avoid waiting, some aggressive drivers attempt to rapidly accelerate in order to pass the intersection before the light changes (De Coensel et al., 2012). Consequently, the combination of high speed and acceleration will inevitably generate more emissions, resulting in this lag in the reduction of PM.

It should be noted that this variation represents the mean values of 30 traffic signal cycles. We know PM variation at intersections is frequently influenced by many random factors, such as instantaneous wind speed, varied wind directions, and traffic-induced turbulence (Tiwary et al., 2011, Wang et al., 2018). It is difficult to capture the stable characteristics of PM variation within one cycle. We selected mean values over 30 repetitions in order to eliminate the influence of random factors and to provide confidence in the values. Furthermore, the vehicles at intersections are allowed to move during the yellow phase; therefore, the three seconds of yellow phase is regarded as a part of the green phase for simplicity.

Fig. 3 shows that during the red-light period, the concentrations of all PM sizes decreased to stable values. This means that such a process could reflect the dispersion pattern of the PM at the intersection. Hence, in order to deeply investigate the dispersion characteristics, the data during the red-light period were collected in this subsection. The mean concentration of PM values are plotted in Fig. 4, and the corresponding box-plots are shown in Fig. S2 (Supplementary Information). The variation of PM in all sizes was found to decrease exponentially. The PM diffused rapidly at the beginning of the red-light period, but there were no longer changes by the end.

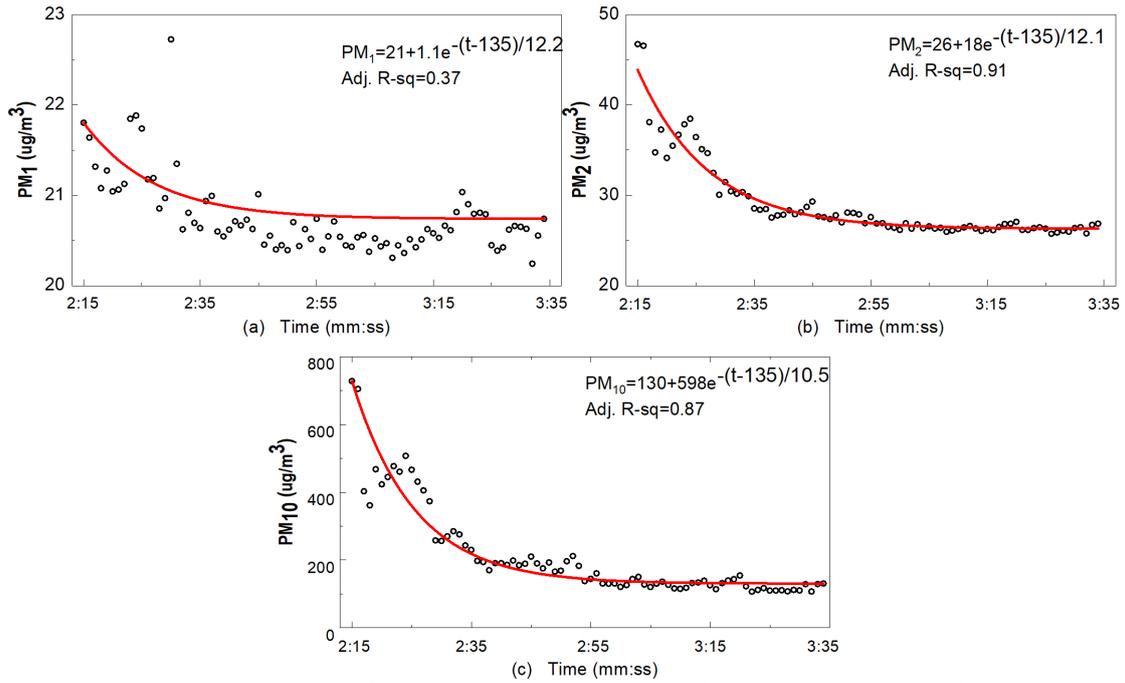

Fig. 4 Mean dispersion of $PM_1$, $PM_2$, and $PM_{10}$ during the red-light period on the sunny day (responding to green status for pedestrian signal).

Additionally, the fitting curves as well as the corresponding mathematical functions were captured individually, which aids our understanding of the exact dispersion of PM at the intersection. In contrast to $PM_{10}$ and $PM_2$, $PM_1$ does not fit well with the negative exponential distribution and exhibits different characteristics in its own magnitude. $PM_1$ demonstrates a small-range variation during the red light period. This means that the dispersion pattern of $PM_1$ is slow. Owing to the small mass and size, $PM_1$ stays aloft and can drift for a long time in the air (Guo et al., 2020).

### 3.2 Variations of PM at the intersection crosswalk through mobile measurements

When the traffic signal turns from green to red (while the pedestrian signal turns from red to green), vehicles are forbidden from moving and the waiting pedestrians begin to use the crosswalk. Pedestrians are inevitably exposed to PM concentrations during their journey (WHO, 2016; Zhang et al., 2020).

In the previous section, we found that PM concentrations exponentially decreased on the roadside during this process. This means that the exposure to these PM concentrations will apparently be different when pedestrians cross the road at varied departure times. In order to examine this, the measured data from six departure journeys (0 s, 5 s, 10 s, 15 s, 20 s, and 30 s.) on the sunny day were collected and are illustrated in Fig. 5. The corresponding box-plots are shown in Fig. S3-S5. From these figures, we observed small changes in $PM_1$ and $PM_2$ concentrations at all departure times. This means that when pedestrians used the crosswalk, they were exposed to nearly identical concentrations of $PM_1$ and $PM_2$. Due to their small size and weight, the submicron particles of $PM_1$ and $PM_2$ will always drift in the air, resulting in stable variations. In contrast, $PM_{10}$ was distinctly different, especially at the beginning of the pedestrian signal phase. Because of their large size and gravity, the coarse particles of $PM_{10}$ fall to the ground and are readily influenced by the turbulence induced by moving pedestrians (Kumar and Goel, 2016; Qiu et al., 2018).

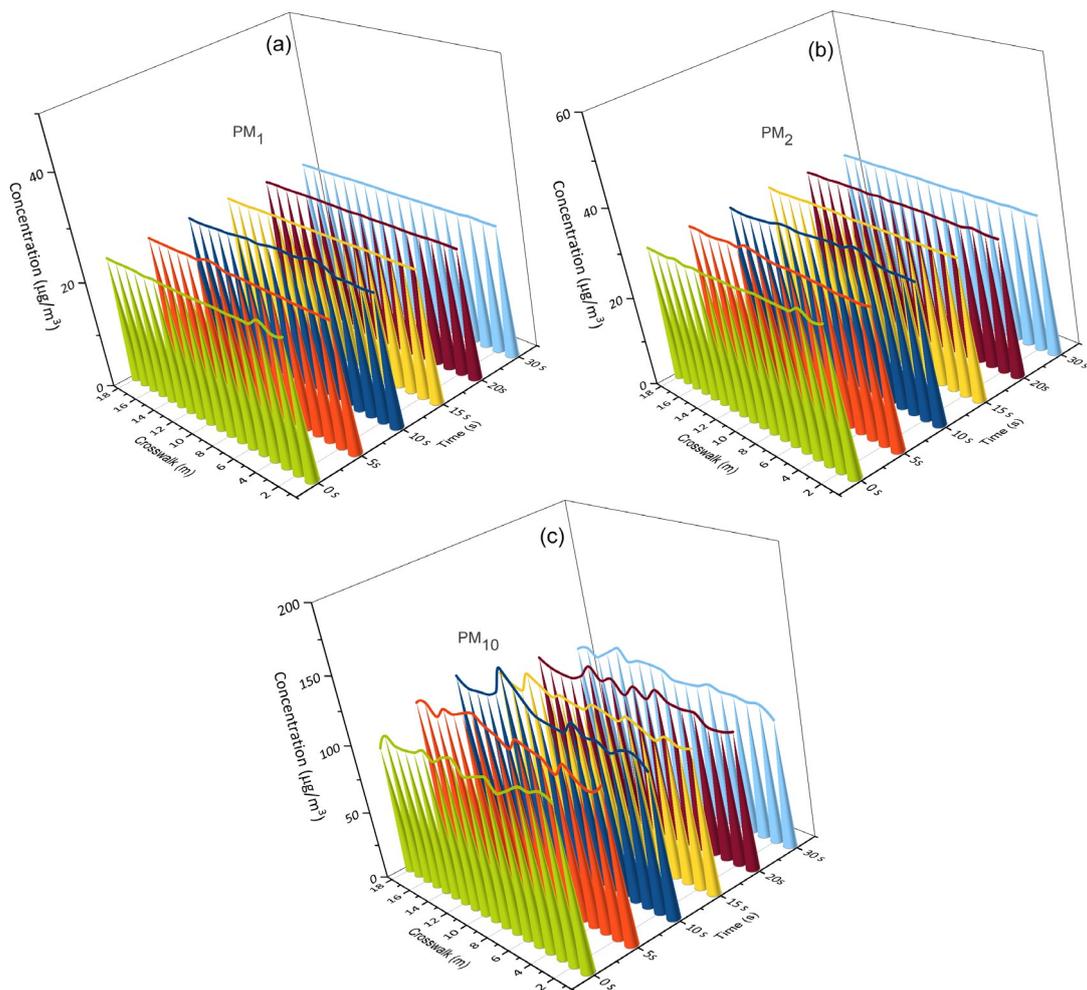

Fig. 5 Mean variations in $PM_1$, $PM_2$, and $PM_{10}$ on the crosswalk at different departure times when pedestrians use the crosswalk on the sunny day.

In addition to the original data presented in Fig. 5, the box plot of the $PM_1$, $PM_2$, and $PM_{10}$ concentrations at different departure times are also displayed in Fig. 6, and reveal the quantitative variations individually. Evidently, the variations of $PM_1$, $PM_2$, and $PM_{10}$ during the first three journeys (0 s, 5 s, 10 s) present an obvious variation, while those in the last three journeys (15 s, 20 s, 30 s) show relatively stable variations. For example, if pedestrians depart as soon as the traffic signal changes (0 s), the varied amplitude of $PM_1$ is between 23.1 and 25.1 µg/m³. In contrast, when pedestrians depart after 15 s, the corresponding amplitude is between 23.0 and 23.2 µg/m³. This shows that with a 15 second delay, the variations of all PM sizes at the crosswalk become smaller and more stable. Consequently, pedestrians could reduce their exposure to PM at the crosswalk by delaying their departure by 15 s.

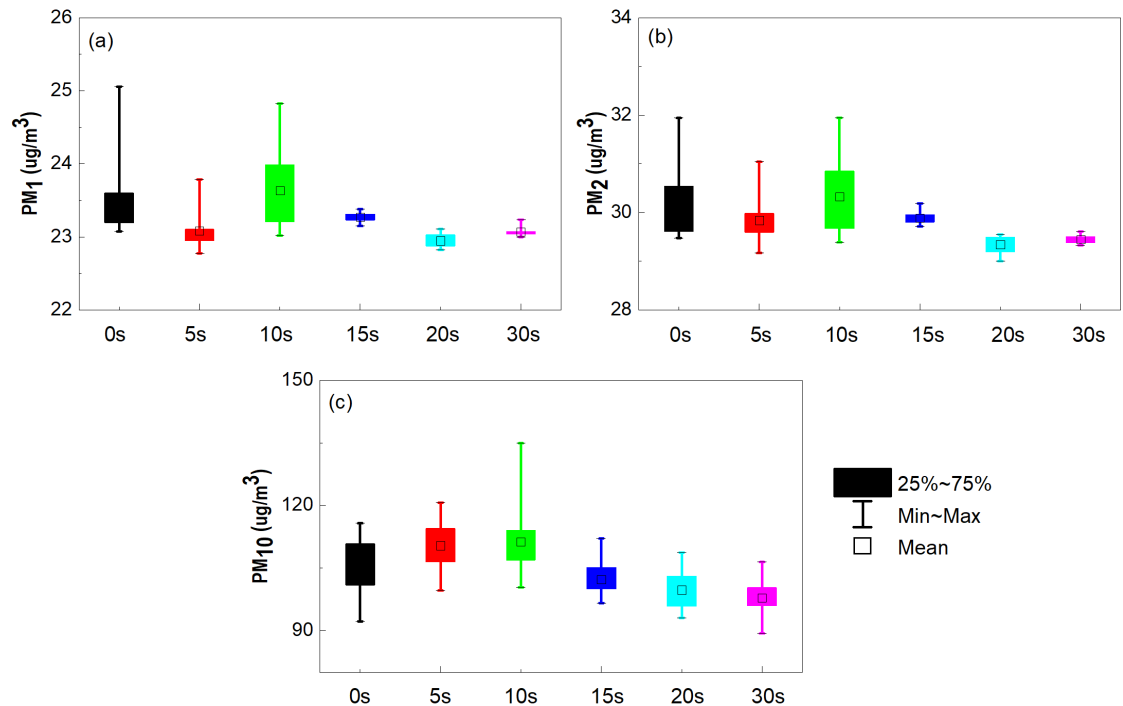

Fig. 6 Box-plots of $PM_1$, $PM_2$, and $PM_{10}$ on the crosswalk during different crossing journeys on the sunny day.

### 3.3 Pedestrian exposure to PM at the intersection

On the basis of the fixed-site and mobile measurements at the intersection, the pollution levels of measured particles are found to be more serious above the ambient levels (WHO, 2006, Qiu et al., 2019; Cao et al., 2020; Zhang et al., 2020). This implies that pedestrians are exposed to high PM concentrations not only on the roadside while they are waiting, but also on the crosswalk while they are crossing. Consequently, it is necessary to adopt respiratory deposition doses (RDD) in order to assess the extent of pedestrian exposure to PM at the intersection.

Considering pedestrian flow at a crosswalk, adults usually cross a crosswalk with a velocity of 1.5 m/s (Marisamynathan and Vedagiri, 2015; Guo et al., 2019).

Consequently, given that the crosswalk in this study spans 16 meters, adults are assumed to breath a total of four times on average during every crossing journey (i.e., 1 s, 4 s, 7 s, 10 s). Owing to these parameters, the RDD values for $PM_1$, $PM_2$, and $PM_{10}$ at the crosswalk during the different journeys were calculated, as shown in Fig. 7.

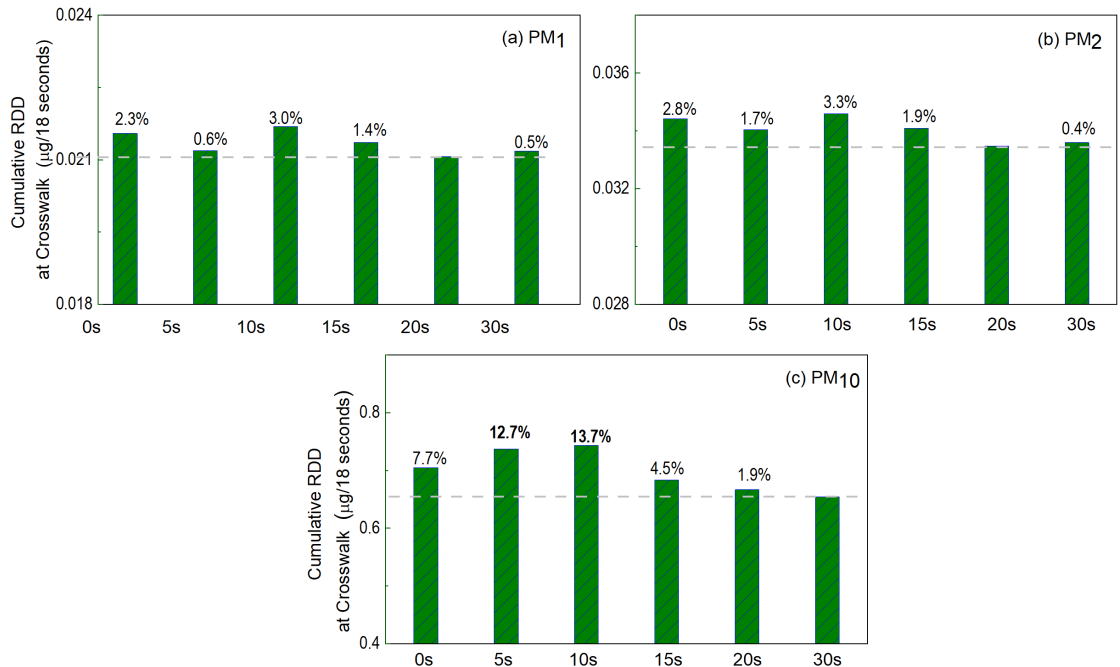

Fig. 7 Cumulative respiratory deposition doses (RDD) of $PM_1$, $PM_2$, and $PM_{10}$ on the crosswalk during different crossing journeys on the sunny day.

Compared to pedestrian exposure during varied journeys, the RDDs of $PM_1$, $PM_2$, and $PM_{10}$ in the first three journeys were all found to be higher than those of the last three journeys. In particular, the maximal RDD for $PM_{10}$ appeared during the third journey and was 13.7% higher than the minimum value (dashed line in Fig 7). The maximal RDD for $PM_2$ and $PM_1$ was found to be 3.3% and 3.0% higher, respectively, than the corresponding minimum. This means that with a delayed departure of 15 s, pedestrians can effectively reduce their exposure to $PM_{10}$ at the crosswalk. However, in reality, some pedestrians always hurry to cross the road as soon as the traffic signal switches, which implies that they are exposed to the highest PM concentrations. Therefore, in order to reduce pedestrian exposure to PM at the crosswalk, it is suggested that pedestrians use the crosswalk only after allowing a time delay.

### 3.4 Comparison of pedestrian exposure to PM in the crosswalk and at the roadside

During the green-light period, pedestrians are exposed to PM on the roadside while waiting. During the red-light period, they are exposed to PM in the crosswalk while crossing. In both situations, pedestrians are ineluctably exposed to PM pollution levels. Therefore, for a better comparison, the RDD for $PM_1$, $PM_2$, and $PM_{10}$ on both the roadside and in the crosswalk are all determined and illustrated in Fig. 8. Furthermore, exposure to PM for both adults and children was taken into account separately. Children are assumed to have crossed the crosswalk at the same velocity of 1.5 m/s, and would therefore breathe five times in total during each journey (i.e., 1 s, 3 s, 5 s, 7 s, 9 s).

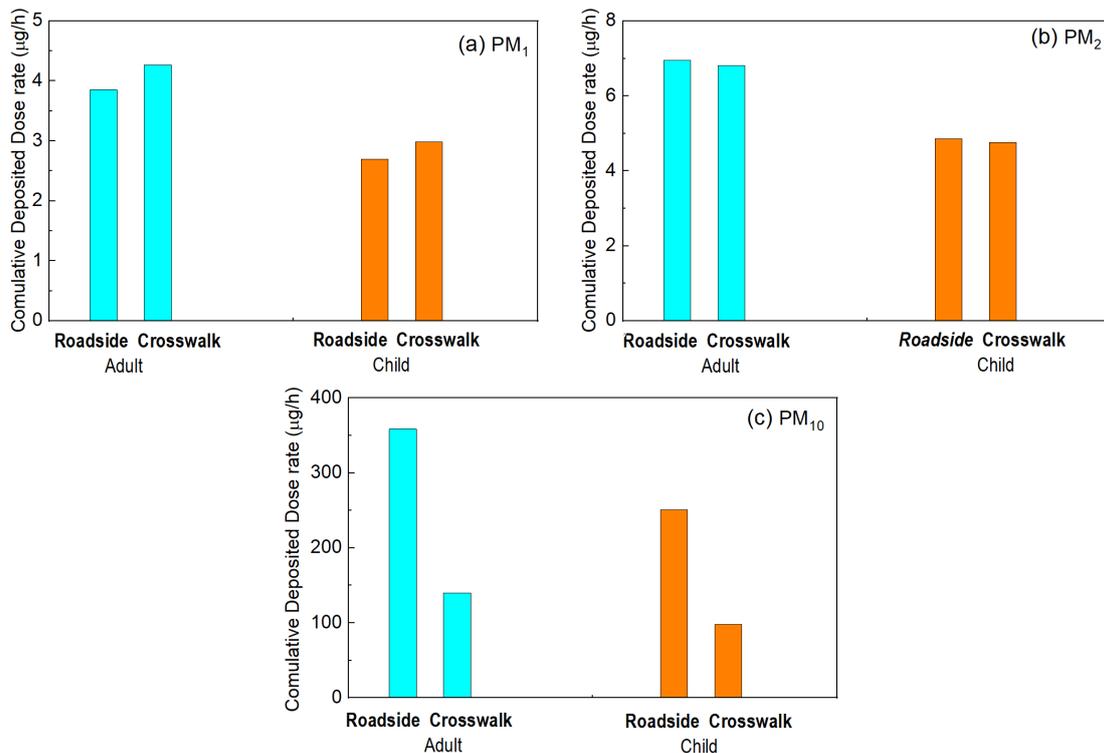

Fig. 8 Comparison of RDD between the roadside while waiting and the crosswalk while crossing at the intersection on the sunny day.

Based on these figures, two points can be addressed. First, pedestrian exposure to $PM_1$ in the crosswalk was found to be slightly higher than that on the roadside; however, pedestrian exposure to $PM_{10}$ in the crosswalk was evidently lower than that on the roadside. In contrast, there is no apparent difference between being in the crosswalk or on the roadside regarding $PM_2$. Second because of the small tidal volume, children's exposure to all sizes of PM was verified as being lower than adult exposure (Guo et al., 2020). However, according to epidemiological evidence, children are more vulnerable to PM and are at a greater health threat, even when exposed to low concentrations. As a result, children are recommended to avoid exposure both on the roadside and in the crosswalk at urban intersections. In summary, although these results are obtained on the basis of limited data, they give rise to the possibility of assessing the risk of being exposed to PM in the crosswalk at the intersection.

### 3.5 Comparison of PM variations on different days and at different intersections

In addition to the results on the sunny day, the measured data on the cloudy day were analyzed in the same manner. The original data on the roadside on the cloudy day also exhibited periodic variation within a traffic signal period (Fig. 9 and Fig. S6), which were in agreement with the results on the sunny day (Fig. 3 and Fig. S1). The only difference was that the particle variations in the two days presented different ranges. This may be caused by the varied background concentration and weather conditions (Tiwary et al., 2011). Similarly, the dispersion characteristics of PM during the red-light period on the cloudy day were investigated, and the PM variations were also confirmed to have exponentially decreased (Fig. 9). Additionally, the pedestrian exposures in crosswalks on the cloudy day were derived in the same manner and displayed in the supplementary file (Fig. S7-S11). In contrast to the results on the sunny day, the pedestrian exposure on the cloudy day was obviously higher on the roadside than in crosswalk.

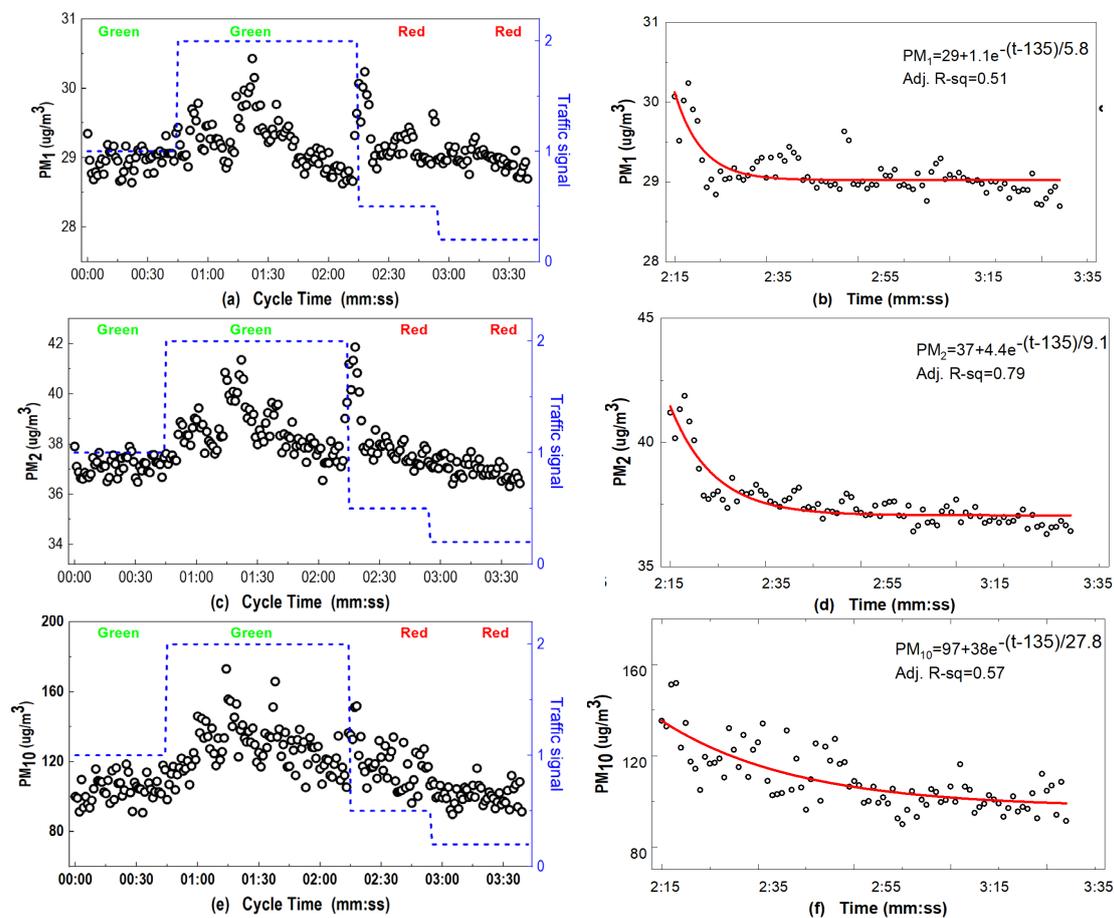

Fig. 9 Mean variation of $PM_1$ (a), $PM_2$ (c), and $PM_{10}$ (e) during the whole cycle time

and mean dispersion of $PM_1$ (b), $PM_2$ (d), and $PM_{10}$ (f) during the red-light period (responding to green status for pedestrian signal) on roadside measurement on the cloudy day. (Traffic index: 1-green traffic light period for left turning vehicles on perpendicular road, 2-green traffic light period for straight vehicles, 0.5-red traffic light period for straight on perpendicular road, 0.2 -red traffic light period for left turning on perpendicular road).

Although the above conclusions are based on two days, they still provide the basic characteristics of PM variation at urban intersections with four phases. In our previous studies, we carried out experiments at an intersection with two phases (He and Lu, 2009; He and Lu, 2012). The results revealed that the variation of particles on the roadside of the intersection was dominated by the traffic signal period and produced periodic variation. Tiwary et al. (2011) regarded these findings as representative of a microscopic view of PM variations at intersections. They cited our entire figure in their review paper and used a separate paragraph for discussion (Tiwary et al., 2011). In this study, we extended the former study to traffic intersections with four phases. In contrast to the former results, the variation of PM at the intersection with four phases becomes more complicated, but still exhibits periodic characteristics. In other words, the results at intersections with four phases as well as results at intersections with two phases provide a microscopic view for understanding the particle variation within the traffic signal period at urban traffic intersections.

## 4 Conclusion

At urban intersections, pedestrians are inevitably exposed to high PM concentrations on the roadside while waiting to cross a road and in the crosswalk while crossing. This study aimed to assess the extent of pedestrian exposure to particles of various sizes at an urban traffic intersection. On the roadside, fixed-site measurements were performed, and particles of various sizes, including $PM_1$, $PM_2$, and $PM_{10}$, all exhibited periodic variations related to the traffic signal cycle. In contrast to submicron particles, the coarse particles and $PM_2$ were found to exhibit exponential distribution during the red-light period. In the crosswalk, mobile measurements at six departure times (i.e., 0 s, 5 s, 10 s, 15 s, 20 s, and 30 s) were performed. The results showed that the exposure levels of all PM sizes in the first three journeys were obviously higher than those during the last three journeys. Finally, based on fixed-site and mobile measurements, we calculated the cumulative deposited doses of PM on the roadside while waiting as well as in the crosswalk while crossing. Pedestrian exposure to submicron particles in the crosswalk was slightly higher than that on the roadside, while exposure to coarse particles showed a pattern of the opposite. While preliminary. Our real-world measurement results contribute to advancing the understanding of pedestrian exposure to PM in crosswalks and assisting the pedestrian to make better informed choice so as to limit PM exposure in these pollution hotspots.

## Acknowledgement

This study was partially funded by the National Natural Science Foundation of China (No. 12072195). The contents of this report reflect the views of the authors, who are responsible for the facts and the accuracy of the information presented herein. This document is disseminated in the interest of information exchange. The 2nd author is funded partially by a grant from the U.S. Department of Transportation's University Transportation Centers Program. However, the U.S. Government assumes no liability for the contents or use thereof.

## References

Adeniran, J.A., Yusuf, R.O., Olajire, A.A. Exposure to coarse and fine particulate matter at and around major intra-urban traffic intersections of Ilorin metropolis, Nigeria. *Atmos. Environ.* 2017, 166: 383-392.

Araji, M.T., Ray, S.D., Leung, L. Pilot-study on airborne $PM_{2.5}$ filtration with particle accelerated collision technology in office environments. *Sustain. Cities Soc.* 2016, 28:101-107.

Brook, R.D., Rajagopalan I.I.I, S.C.A.P., Brook, J.R., Bhatnagar, A., Diez-Roux, A.V., Holguin, F., Hong, Y., Luepker, R.V., Mittleman, M.A., Peters, A., Siscovick, D., Smith, S.C., Whitsel, L., Kaufman, J.D. Particulate matter air pollution and cardiovascular disease an update to the scientific statement from the American Heart Association. *Circulation* 2010, 121:2331-2378.

Cao, R., Li, B., Wang, Z.Y., Peng, Z. R, Tao, S.K., Lou, S.R. Using a distributed air sensor network to investigate the spatiotemporal patterns of $PM_{2.5}$ concentrations. *Environ. Pollut.* 2020, 264: 114549.

De Coensel, B., Can, A., Degraeuwe, B., De Vlieger, I., Botteldooren, D. Effects of traffic signal coordination on noise and air pollutant emissions. *Environ. Model. Softw* 2012, 35: 74-83.

De Nazelle, A., Rodriguez, D.A., Crawford-Brown, D. The built environment and health: Impacts of pedestrian-friendly designs on air pollution exposure. *Sci. Total Environ.* 2009, 407: 2525-2535.

Fann, N., Baker, K.R., Fulcher, C.M. Characterizing the $PM_{2.5}$-related health benefits of emission reductions for 17 industrial, area and mobile emission sectors across the U.S. *Environ Int.* 2012, 49: 141-151.

Guo, L.L., Johnson, G.R., Hofmann, W., Wang, H., Morawska, L. Deposition of ambient ultrafine particles in the respiratory tract of children: A novel experimental method and its application. *J. Aerosol Sci.* 2020, 139: 105465.

Goel, A and Kumar, P. Vertical and horizontal variability in airborne nanoparticles and their exposure around signalised traffic intersections. *Environ. Pollut.* 2016, 214: 54-69.

Gokhale, S., Raokhande, N. Performance evaluation of air quality models for predicting $PM_{10}$ and $PM_{2.5}$ concentrations at urban traffic intersection during winter period. *Sci. Total Environ.* 2008, 394, 9-24.

Hao, C.R., Xie, X.M., Huang, Y., Huang, Z. Study on influence of viaduct and noise barriers on the particulate matter dispersion in street canyons by CFD modeling. *Atmos. Pollut. Res.* 2019, 10: 1723-1735.

He, H.D., Lu, W.Z. Urban aerosol particulates on Hong Kong roadsides: size distribution and concentration levels with time. *Stoch Environ Res Risk Assess.* 2012, 26: 177-187.

He, H.D., Lu, W. Z. Measurement of particulate matter at intersection in Hong Kong. The Seventh International Conference on Urban Climate, Yokohama, Japan 29 Jun–3 July, 2009.

Heo, J., Adams, P.J., Gao, H.O., Public health costs accounting of inorganic $PM_{2.5}$ pollution in metropolitan areas of the United States using a risk-based source-receptor model. *Environ Int.* 2017, 106: 119-126.

Hinds, W. C. Aerosol Technology: Properties, Behaviour, and Measurement of Airborne Particles, Wiley Interscience, 2nd edn, 1999, p. 483.

Huang, R.J., Zhang, Y.L., Bozzetti, C., Ho, K.F., Cao, J.J., Han, Y.M., Daellenbach, K.R., Slowik, J.G., Platt, S.M., Canonaco, F., et al. High secondary aerosol contribution to particulate pollution during haze events in China. *Nature* 2014, 514: 218-222.

Ishaque, M.M., Noland, R.B. Simulated pedestrian travel and exposure to vehicle emissions. *Transp. Res. D.* 2018, 13: 27-46.

Khan, M.F., Hamid, A.H., Bari, M.A., Tajudin, A.B.A., Latif, M.T., Nadzir, M.S.M., Sahani, M., Wahab, M.I.A., Yusup, Y., Maulud, K.N., Yusoff, M.F., Amin, N., Akhtaruzzaman, M., Kindzierski, W., Kumar, P. Airborne particles in the city center of Kuala Lumpur: Origin, potential driving factors, and deposition flux in human respiratory airways. *Sci. Total Environ.* 2019, 650:1195-1206.

Kumar, P., Goel, A. Concentration dynamics of coarse and fine particulate matter at and around the signalised traffic intersections. *Environ Sci Process Impacts.* 2016, 18:1220-1235.

Landrigan, P.J., Fuller, R., Fisher, S., Suk, W.A., Sly, P., Chiles, T.C., Bose-O'Reilly, S. Pollution and children's health. *Sci. Total Environ.* 2019, 650: 2389-2394.

Li, Q.F., Wang, Z.A., Yang, J.G., Wang, J.M. Pedestrian delay estimation at signalized intersections in developing cities. *Transp. Res. A.* 2005, 39: 61-73.

Liu, H.B., Wang, W., Chen, K., Yin, Y., Zheng, H., Wu, J., Qin, S. Liu, J.H., Feng, Y.K., Yan, Y.Y., Liu, D.T., Zhao, D.L., Qi, S.H. Size-segregated carbonaceous aerosols emission from typical vehicles and potential depositions in the human respiratory system. *Environ. Pollut.* 2020, 264: 114705.

Liu, Y.Q. Pedestrians' speed analysis for two-stage crossing at a signalized intersection. *Civ. Eng. J.* 2019, 5: 505-514.

Marisamynathan, S., Vedagiri, P. A statistical analysis of pedestrian behaviour at signalized intersections. *European Transport\Trasporti Europei* 2015 57:7.

Matus, K., Nam, K.M., Selin, N.E., Lamsal, L.N., Reilly, J.M., Paltsev, S. Health damages from air pollution in China. *Glob Environ Change.* 2012, 22: 55-66.

Pan, S., Roy, A., Choi, Y.S., Sun, S.Q., Gao, H.O. The air quality and health impacts of projected long-haul truck and rail freight transportation in the United States in 2050. *Environ Int.* 2019, 130:104922.

Qiu, Z.W., Song, J.H., Hao, C.H., Li, X.X., Gao, H.O. Investigating traffic-related PM exposure on and under pedestrian bridges: A case study in Xi'an, China. *Atmos. Pollut. Res.* 2018, 9: 877-886.

Qiu, Z.W., Wang, W.Z., Zheng, J.L., Lv, H.T. Exposure assessment of cyclists to UFP

and PM on urban routes in Xi'an, China. *Environ. Pollut.* 2019, 250: 241-250.

Quiros, D.C., Lee, E.S., Wang, R., Zhu, Y. Ultrafine particle exposure while walking, cycling, and driving along an urban residential roadway. *Atmos. Environ.* 2013, 73:185-194.

Sadeghi, B., Choi, Y.S., Yoon, S., Flynn, J., Kotsakis, A., Lee, S.J. The characterization of fine particulate matter downwind of Houston: Using integrated factor analysis to identify anthropogenic and natural sources. *Environ. Pollut.* 2020, 262: 114345.

Sanchez-Soberon, F., Mari, M., Kumar, V., Rovira, J., Nadal, M., Schuhmacher, M. An approach to assess the particulate matter exposure for the population living around a cement plant: modelling indoor air and particle deposition in the respiratory tract. *Environ. Res.* 2015, 143:10-18.

Scungio, M., Stabile, L., Rizza, V., Pacitto, A., Russi, A., Buonanno, G. Lung cancer risk assessment due to traffic-generated particles exposure in urban street canyons: A numerical modelling approach. *Sci. Total Environ.* 2018, 631: 1109-1116.

Tittarelli, A., Borgini, A., Bertoldi, M., De Saeger, E., Ruprecht, A., Stefanoni, R., Tagliabue, G., Contiero, R., Crosignani, P. Estimation of particle mass concentration in ambient air using a particle counter. *Atmos. Environ.* 2008, 42: 8543-8548.

Tiwary, A., Robins, A., Namdeo, A., Bell, M. Air flow and concentration fields at urban road intersections for improved understanding of personal exposure. *Environ Int.* 2011, 37: 1005-1018.

Tuch, T., Mirme, A., Tamm, E., Heinrich, J., Heyder, J., Brand, P., Roth, C., Wichmann, H.E., Pekkanen, J., Kreyling, W.G. Comparison of two particle size spectrometers from ambient aerosol measurements in environmental epidemiology. *Atmos. Environ.* 2000, 34:139-49.

Wang, X., Gao, H.O. Exposure to fine particle mass and number concentrations in urban transportation environments of New York City. *Transp. Res. D.* 2011, 16:384-391.

Wang, Z.Y., Zhong, S.Q., He, H.D., Peng, Z.R., Cai, M. Fine-scale variations in $PM_{2.5}$ and black carbon concentrations and corresponding influential factors at an urban road intersection. *Build Environ.* 2018, 141:215-225.

Wittmaack, K. Advanced evaluation of size-differential distributions of aerosol particles. *J. Aerosol Sci.* 2002, 33:1009-1025.

World Health Organization. WHO Air quality guidelines for particulate matter, ozone, nitrogen dioxide and sulfur dioxide. 2016, Geneva, Switzerland.

WHO/SDE/PHE/OEH/06.02.

Zhai, S.R., Albritton, D. Airborne particles from cooking oils: Emission test and analysis on chemical and health implications. *Sustain. Cities Soc.* 2020, 52: 101845.

Zhang, L., Zhang, Z.Q., McNulty, S., Wang, P. The mitigation strategy of automobile generated fine particle pollutants by applying vegetation configuration in a street-canyon. *J. Clean. Prod.* 2020, 17: 122941.